\documentstyle{laa}
\input psfig
\begin{document}
\def\Boxtimes{\protect\pBoxtimes}
\def\pBoxtimes{{\ooalign{\hfil\raise.2ex\relax
\hbox{$\times$}\hfil\crcr\hbox{$\Box$}}}}
\def\Boxcirc{\protect\pBoxcirc}
\def\pBoxcirc{{\ooalign{\hfil\raise.2ex\relax
\hbox{$\circ$}\hfil\crcr\hbox{$\Box$}}}}
\def\Diatimes{\protect\pDiatimes}
\def\pDiatimes{{\ooalign{\hfil\raise.2ex\relax
\hbox{$\times$}\hfil\crcr\hbox{$\Diamond$}}}}
\def\Boxplus{\protect\pBoxplus}
\def\pBoxplus{{\ooalign{\hfil\raise.2ex\relax
\hbox{$+$}\hfil\crcr\hbox{$\Box$}}}}
%
%
\newcommand{\araa}{ARA\&A}   
\newcommand{\afz}{Afz}       
\newcommand{\aj}{AJ}         
\newcommand{\azh}{AZh}       
\newcommand{\aaa}{A\&A}      
\newcommand{\aas}{A\&AS}     
\newcommand{\aar}{A\&AR}     
\newcommand{\apj}{ApJ}       
\newcommand{\apjs}{ApJS}     
\newcommand{\apjl}{ApJL}     
\newcommand{\apss}{Ap\&SS}   
\newcommand{\baas}{BAAS}     
\newcommand{\jaa}{JA\&A}     
\newcommand{\mnras}{MNRAS}   
\newcommand{\nat}{Nat}       
\newcommand{\pasj}{PASJ}     
\newcommand{\pasp}{PASP}     
\newcommand{\paspc}{PASPC}   
\newcommand{\qjras}{QJRAS}   
\newcommand{\sci}{Sci}       
\newcommand{\sova}{SvA}      
\newcommand{\rasav}{Ric.~astr.~Specola astr.~Vatic.} 
\newcommand{\sca}{Scient.~Am.}   
\newcommand{\stel}{Sky Telesc.}  
\newcommand{\spsrev}{Space Sci.~Rev.} 
\newcommand{\phfl}{Phys. Fluids} 
\newcommand{\phrev}{Phys. Rev.} 
\newcommand{\rprph}{Rep. Prog. Phys.} 
\newcommand{\rmph}{Rev. Mod. Phys.} 
\newcommand{\jplph}{J. Plasma Phys.} 
\newcommand{\jmph}{J. Math. Phys.} 
\newcommand{\jgeores}{J. Geophys. Res.}
\newenvironment{refs}{\bigskip 
   \begin{flushleft} {\Large\bf References} \end{flushleft} \bigskip 
   \begin{list}{}{\setlength{\leftmargin}{1cm}\setlength{\itemindent}{-1cm}
   \setlength{\topsep}{0cm}\setlength{\itemsep}{-0.12cm}} 
   \vspace*{-0.6cm}}{\end{list} }
\newenvironment{tabrefs}{\bigskip 
   \begin{list}{}{\setlength{\leftmargin}{1cm}\setlength{\itemindent}{-1cm}
   \setlength{\topsep}{0cm}\setlength{\itemsep}{-0.2cm}} 
   \vspace*{-0.6cm}}{\end{list}}
\hyphenation {non-re-la-ti-vi-stic}
%
%
%
\newcommand{\cm}{\,{\rm cm}}
\newcommand{\mG}{\,{\rm mG}}
\newcommand{\dyn}{\,{\rm dyn}}
\newcommand{\erg}{\,{\rm erg}}
\newcommand{\kpc}{\,{\rm kpc}}
\newcommand{\yr}{\,{\rm yr}}
\newcommand{\secnd}{\,{\rm s}}
\newcommand{\Msun}{\,\mbox{M}_{\odot}}
\newcommand{\smm}[1]{\mbox{\small\rm #1}}
\newcommand{\footm}[1]{\mbox{\footnotesize\rm #1}}
\newcommand{\scrm}[1]{\mbox{\scriptsize\rm #1}}
\newcommand{\tinm}[1]{\mbox{\tiny\rm #1}}
\newcommand{\proptosim}{\, \raisebox{-1.1mm}{\mbox{\scriptsize 
                    $\stackrel{\propto}{\sim}$}}\,}
\def\lmean{\mathopen{<}}
\def\rmean{\mathclose{>}}
%
%
\newcommand{\as}[2]{$#1''\hspace{-1.8mm}.\hspace{.4mm}#2$}
\newcommand{\am}[2]{$#1'\hspace{-1.1mm}.\hspace{.1mm}#2$}
\newcommand{\m}[2]{$#1^{m}\hspace{-1.8mm}.\hspace{.4mm}#2$} 
\newcommand{\days}[2]{$#1^{\scrm{d}}\hspace{-1.8mm}.\hspace{.4mm}#2$}
\newcommand{\etal}{et al.\ }
\newcommand{\etalk}{et al.,\ }
\newcommand{\ie}{i.e.,\ }
\newcommand{\eg}{e.g.,\ }
\newcommand{\yes}{{\tt *}}
\newcommand{\gtsim}{$\, \raisebox{-1.1mm}{\scriptsize $\stackrel{>}{\sim}$}
                    \,$}
\newcommand{\ltsim}{$\, \raisebox{-1.1mm}{\scriptsize $\stackrel{<}{\sim}$}
                    \,$}
\newcommand{\bm}[1]{\mbox{{\protect{\boldmath #1}}}} 
\newcommand{\Mzon}{M$_{\odot}$}
\newcommand{\Lzon}{L$_{\odot}$}
\newcommand{\Td}{$T_{\scrm{d}}$}
\newcommand{\Md}{$M_{\scrm{d}}$}
\newcommand{\MdI}{$M_{\scrm{d,\,{\sc iras}}}$}
\newcommand{\MdO}{$M_{\scrm{d,\,opt}}$}
\newcommand{\Mg}{$M_{\scrm{gas}}$}
\newcommand{\Mav}{$M_{{\scrm{d,}}\,A_V}$}
\newcommand{\Mebv}{$M_{{\scrm{d,}}\,E_{B-V}}$}
\newcommand{\MHI}{$M_{\mbox{\scriptsize\sc H$\,$I}}$}
\newcommand{\Rv}{$R_{\scrm{V}}$}
\newcommand{\Ab}{$A_{\scrm{B}}$}
\newcommand{\Av}{$A_{\scrm{V}}$}
\newcommand{\Ai}{$A_{\scrm{I}}$}
\newcommand{\EBV}{$E_{B-V}$}
\newcommand{\EVI}{$E_{V-I}$}
\newcommand{\EBI}{$E_{B-I}$}
\newcommand{\kms}{\mbox{km s$^{-1}$}}
\newcommand{\ergs}{\mbox{erg s$^{-1}$}}
\newcommand{\ergcms}{erg s$^{-1}$ cm$^{-2}$} 
\newcommand{\Mpc}{Mpc$^{-1}$} 
\newcommand{\Ha}{H$\alpha$}
\newcommand{\Hb}{H$\beta$}
\newcommand{\Hg}{H$\gamma$}
\newcommand{\lda}{$\lambda$}
\newcommand{\OI}{[{\sc O$\,$i}]}
\newcommand{\OII}{[{\sc O$\,$ii}]}
\newcommand{\OIII}{[{\sc O$\,$iii}]}
\newcommand{\NI}{[{\sc N$\,$i}]}
\newcommand{\NII}{[{\sc N$\,$ii}]}
\newcommand{\NeIII}{[{\sc Ne$\,$iii}]}
\newcommand{\SII}{[{\sc S$\,$ii}]}
\newcommand{\HI}{{\sc H$\,$i}}
\newcommand{\HII}{{\sc H$\,$ii}}
\newcommand{\um}{\mbox{\rm $\mu$m}}
\newcommand{\plm}{$\, \pm \,$}
\newcommand{\upp}{$<\;$}
\newcommand{\low}{$>\;$}
\newcommand{\multi}{\multicolumn}
\newcommand{\fnsize}{\footnotesize}
\newcommand{\sub}[1]{$_{\mbox{\scriptsize #1}}$}
\newcommand{\subsl}[1]{$_{\mbox{\scriptsize\sl #1}}$}
\newcommand{\super}[1]{$^{\mbox{\scriptsize #1}}$}
\newcommand{\quarter}{\frac{1}{4}}
\newcommand{\half}{\frac{1}{2}}
\newcommand{\sep}[1]{@{\extracolsep{#1}}}

\thesaurus{09.11.1; 11.09.1 NGC~128;11.09.2; 11.11.1; 11.14.1}

\title{A counter-rotating tilted gas disc in the peanut galaxy NGC~128
\thanks{Based on observations taken with the Canada-France-Hawaii
Telescope, operated by the National Research Council of Canada, the Centre National de la
Recherche Scientifique of France, and the University of Hawaii}}

\author{Eric Emsellem \inst{1} and Robin Arsenault \inst{2}}

\offprints{E. Emsellem (email: eemselle@eso.org)}

\institute{European Southern Observatory, Karl-Schwarschild Strasse 2, D-85748
Garching b. M\"unchen, Germany \and
Canada-France-Hawaii Corporation, P.O. Box 1597, Kamuela, HI 96743, USA}

\date{accepted 07/01, 1996}

\maketitle
\markboth{E. Emsellem \& R. Arsenault: A counter-rotating tilted gas disc in NGC~128}{}

\begin{abstract}
We have obtained $V$, $R_c$, $I_c$ HRCAM images and TIGER 
spectrography of the central region of the peanut galaxy NGC~128.
The colour images reveal the presence of a red disc tilted by about 26 degres
with respect to the major-axis of the galaxy. This tilted disc is made of
dust and gas, as revealed by the 2D TIGER map of the ionized gas distribution. 
The TIGER stellar and gas velocity
fields show that the angular momentum vectors of the stellar and gaseous components
are reversed. We therefore suggest that the gas orbits belong to the 
so-called anomalous family, which is evidence for a tumbling triaxial
potential (a bar) associated with the peanut morphology. The bar formation has very probably
been triggered through the interaction with its nearby companion NGC~127, from which
the dissipative component is being accreted along retrograde orbits.
\end{abstract}

\section{Introduction}

N body simulations have shown that strong bars can appear 
as peanut-shaped bulges when viewed close to edge-on
(Combes \& Sanders 1981, Combes et al.  1990, Pfenniger \& Friedli 1991).
The same bar would be seen nearly round when end-on and 
box-shaped for an intermediate viewing angle. These morphologies are indeed
observed in galaxies, which suggested that boxy and peanut-like bulges
could be linked to the presence of bars. Recently, Kuijken \& Merrifield
(1995) have shown that it is possible to detect edge-on bars kinematically
by a careful study of the projected velocity distribution. When 
a tumbling triaxial structure is present, the observed
Line Of Sight Velocity Distributions (LOSVDs) should exhibit contrasted gaps 
corresponding to the transition regions between the main resonances.
They indeed detected these expected features in two peanut-shaped galaxies.
A larger sample has recently been analysed by Bureau \& Freeman 
(1997) who found the signature of a bar in nearly all systems.
This technique is powerful to detect bars in edge-on systems.
However it requires the presence of a rather extended gas 
disc\footnote{Since star orbits can ``cross'' each other,
the bar signature is weaker in stellar LOSVDs (see Kuijken \& Merrifield\ 1995).}. It may therefore
be a difficult task to apply this method systematically on all boxy/peanut bulges.

In this Letter, we show that in exceptional circumstances the signature of the bar
is even more obvious. This is illustrated with the
prototype of peanut-shaped galaxy, namely NGC~128.
Although this ``peculiar'' S0 galaxy appears in the Hubble Atlas, it has been
only scarcely studied. Bertola \& Capaccioli (1977) published a combined
photometric and spectroscopic analysis of NGC~128, including the derivation of
its major-axis velocity profile. It was also included in the sample of 
early-type galaxies observed by Bertola et al. (1992) who obtained
long-slit spectrography along the major-axis, and only commented on the
mean angular momenta of the stars and gas concluding that they had the same direction.
This is, as will be shown in this Letter,
inconsistent with our TIGER data (preliminary results published in
Monnet et al. 1995; see also Pagan et al. 1996 and Kuijken et al. 1996).

Details on our observations are given in Sect.~\ref{sec:obs}.
The corresponding results are presented in Sect.~\ref{sec:res}. A brief
discussion and some conclusions are drawn in Sect.~\ref{sec:conc}.

\section{Observations}
\label{sec:obs}

\subsection{Photometry}

$V$, $R_c$ and $I_c$ (Cousin system) images of NGC~128 were obtained
with HRCAM at the CFHT in Dec. 1993. These exposures were reduced 
in the classical way under IRAF. Bad columns present on the CCD were 
interpolated using the adjacent pixels. We  normalized the 
images using the available aperture photometry 
(de Vaucouleurs \& Longo 1988; Poulain 1986, 1988), leading to
an uncertainty of about 0.05 magnitude. All images were then centred and rotated
such that the major-axis (for $R < 10\arcsec$) of NGC~128 is horizontal
(measured PA of $1.9 \pm 1$  degres). The resulting spatial resolution is poor and nearly identical 
for all images: FWHM of $\sim 1\farcs1$.
We computed the $V - R_c$ and $V - I_c$ colour images, which were 
smoothed through a simple gaussian convolution with $\sigma = 0\farcs25$. 

\subsection{Spectroscopy}

We also observed NGC~128 in Nov.~1993 using 
the integral field spectrograph TIGER mounted on the
Cassegrain focus of the CFHT. The lens diameter was set to $0\farcs39$ giving
a field of $\sim 7\arcsec \times 7\arcsec$. Two spectral domains 
were covered: one around the Mg triplet (5100--5500~\AA) and 
another including the \NII/\Ha\ and \SII\ emission lines (6530--6990~\AA)
both with a sampling of $1.5$\AA/px. The standard reduction steps were 
applied to extract and calibrate the spectra of each individual exposure, using the 
TIGER software (Bacon et al. 1994). A total of 3 hours were obtained for the
absorption line domain ($\sigma_{\star} = 0\farcs47$, 
$\sigma_{spec} \sim 75$~km.s$^{-1}$; hereafter ``blue spectra'') 
and 2.5 hours for the emission lines ($\sigma_{\star} =  0\farcs42$, 
$\sigma_{spec} \sim 60$~km.s$^{-1}$; hereafter ``red spectra'')

The stellar kinematics were derived from the blue spectra using the FCQ method of Bender (1990)
which provides the full LOSVD for each spatial
element. We will however only report here the estimates of the two first moments
derived from a gaussian fit: at our spectral (and spatial)
resolution the derived LOSVDs in the central part of NGC~128
do not deviate significantly from single Gaussians.

The red spectra of NGC~128 exhibit the \Ha, \NII\ and \SII\ emission lines,
all detected in the central part of NGC~128. In order to obtain pure 
emission line spectra, we first had to subtract the stellar
continuum from the original spectra. We thus used
a high resolution spectrum of the emission-line free elliptical NGC~596 
kindly provided by Paul Goudfrooij.
The relative doppler shift $\Delta V$ and broadening $\Delta \sigma$ 
of the blue part of NGC~596's spectrum with respect to the TIGER blue spectra 
of NGC~128 were determined using FCQ.
The broadened spectrum of NGC~596 was then scaled to each individual
TIGER red spectrum of NGC~128 and subtracted. Low frequency differences between the
stellar continua of NGC~596 and NGC~128 were included using a low-order polynomial.
The resulting pure emission line spectra were analysed with the
FITSPEC software written by A. Rousset (Lyon Observatory).

\section{Results}
\label{sec:res}

\subsection{The morphology of NGC~128}

\begin{figure}
\psfig{figure=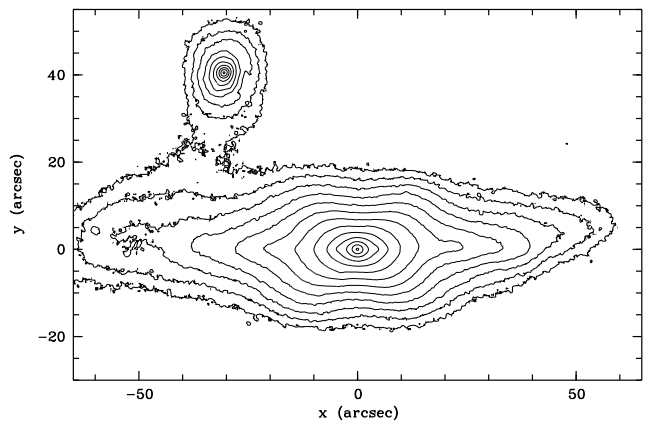,width=8.8cm}
\caption[]{HRCAM $I_c$ band image of the prototypical peanut galaxy
NGC~128. The step is 0.5 mag.arcsec$^{-2}$ and the faintest isophote
corresponds to 21.5 mag.arcsec$^{-2}$. The small galaxy in the top left is NGC~127, 
a small interacting companion. The interaction between the two galaxies 
is clearly seen here.}
\label{fig:hrc}
\end{figure}

As already emphasized in previous studies (e.g. Bertola \& Capaccioli 1977),
the outer disc of NGC~128 is bending towards its companion NGC~127 (westwards),
certainly due to the interaction (Fig.~\ref{fig:hrc}). 
The NGC~128 group contains another small galaxy NGC~130:
although the latter probably participates in the interaction
(Jarvis 1990), we did not detect any luminous bridge between NGC~128 and NGC~130.
The peanut in NGC~128 appears as an additional ``quadrupole
component'' whose maxima lie at $x = \pm 10\arcsec$ (along the major-axis) 
and $y = \pm 8\arcsec$ (minor-axis). NGC~128 also 
contains a central thin disc extending up to $\sim 7\arcsec$. 

\subsection{A tilted red disc}

\begin{figure}
\psfig{figure=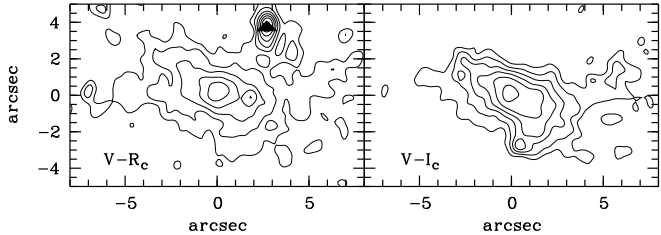,width=8.8cm}
\caption[]{$V - R_c$ (left) and $V - I_c$ (right) isocolours of the central region of
NGC~128: the step is 0.01 magnitude and the maxima are 0.58 and 1.34 respectively for
the $V - R_c$ and $V - I_c$ contours. The secondary maximum in the $V - R_c$ plot
is due to a CCD defect (black triangle).}
\label{fig:col}
\end{figure}

In the central $6\arcsec$, the $V - R_c$ and $V - I_c$ colour 
images exhibit flattened isocontours which
are tilted by $\sim 26$ degres with respect to the major-axis of the galaxy (Fig.~\ref{fig:col}).
The $R_c$ band includes the \Ha, \NII\ and \SII\ emission lines
which could produce part of this reddening. However, our $I_c$ filter only
contains the [SIII]$\lambda$9069 emission line which should represent
less than $10$\% of the total nebular emission in the $R_c$ filter.
The reddening in $V - I_c$ would correspond to a relative extinction of E$(V - I_c) \sim 0.02$.
In order to search for the presence of dust we have symmetrized the $V$ band 
image with respect to the minor-axis: we found a difference of $\sim 0.045$ magnitude
between the surface brightness along the disc and at its symmetric point. This
would be consistent with the $V - I_c$ reddening being entirely due to dust extinction
(assuming ``Galactic dust'', Emsellem 1995). The expected dust reddening E$(V - R_c)$
would then be smaller than 0.01 magnitude.

\subsection{Gas distribution and kinematics}

\begin{figure}
\psfig{figure=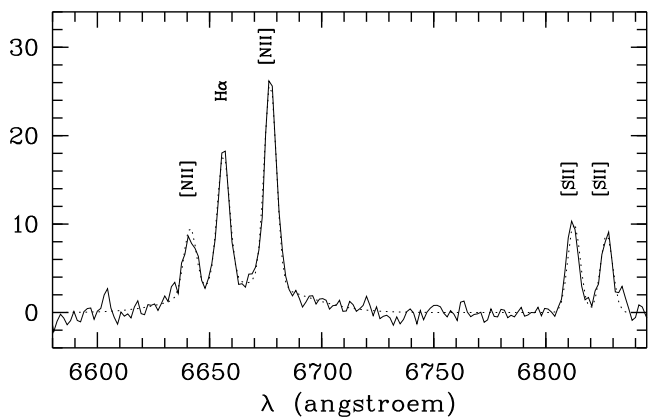,width=8.8cm}
\caption[]{Sum of the 5 central TIGER spectra (stellar continuum subtracted)
showing the \NII, \Ha, and \SII\ narrow lines as well as the broad \Ha\ line.
The best fit derived with FITSPEC is superimposed (dotted line).}
\label{fig:BLR}
\end{figure}

The red spectra exhibit rather narrow emission lines ($85 < \sigma \mbox{(km.s$^{-1}$)} < 135$)
throughout the TIGER field except in the central arcsecond where a superimposed broad H$\alpha$ 
component is marginally detected. This component seems
to be redshifted with respect to the narrow line system by $\sim 450$km.s$^{-1}$, and
has a FWHM of $\sim 2590$~km.s$^{-1}$ (Fig.~\ref{fig:BLR}). However, the presence
of a Broad Line Region (BLR) should be confirmed with higher signal to noise data.

\begin{figure}
\psfig{figure=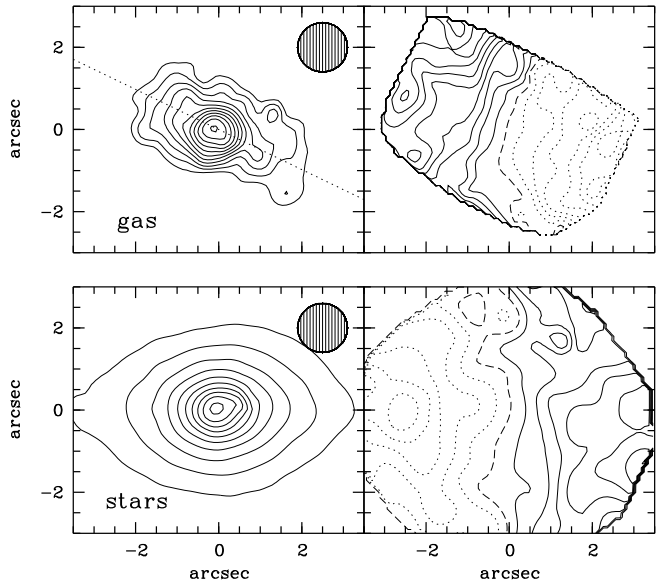,width=8.8cm}
\caption[]{2D distributions (left panels) and velocity fields (right panels)
of the gas (\NII\lda6583 line, top) and stellar (bottom) components.
The dotted straight line in the top left panel shows the gas mean major-axis at
26~degres. The isovelocity step is 25~km.s$^{-1}$ in both cases (the dotted lines
correspond to negative velocities, the solid lines to positive). The 
gas maps have been smoothed by a gaussian with $\sigma = 0\farcs3$. The
drawn beams correspond to the seeing FWHMs in each configuration.}
\label{fig:stargas}
\end{figure}

Fig~\ref{fig:stargas} shows the gas distribution of the \NII\lda6583 emission line:
the gas is distributed in a disc-like structure 
whose major-axis is tilted by $\sim 26$ degres with respect 
to the $I_c$ band major-axis of the core.
This is very similar to the central feature 
detected in the $V - R_c$ image which demonstrates that the gas disc
certainly extends at least up to $6\arcsec$ ($\sim 1750$~pc at 60 Mpc)
from the centre. The \NII\lda6583 to \Ha\ ratio 
slightly rises towards the centre from about 1.2 to 1.65. The ratios \SII$/$\Ha\ 
and \SII\lda6717$/$\SII\lda6731 are in the ranges $[0.8-1.5]$ and $[0.7-1.5]$
respectively, typical for a LINER, and consistent with the gas being 
photoionized by post-AGB stars (Binette et al. 1994). 

The TIGER gas velocity field beautifully confirms the tilt of the gas disc:
the zero isovelocity (w.r.t. the systemic velocity of the galaxy) is tilted by
$\sim 25$ degres consistent with the position angle of the gas distribution minor-axis.
Since dust is associated with the gaseous component,
we rejected the hypothesis of a linear ejection from the centre.
At our resolution, the gas thus exhibits nearly cylindrical rotation 
with $\Omega \sim 67$~km.s$^{-1}$.arcsec$^{-1}$. 

\subsection{Stellar kinematics}

The picture gets even more striking when we compare the velocity fields of
the stellar and gaseous components (Fig.~\ref{fig:stargas}): their mean 
angular momenta are reversed. The maximum stellar velocities ($\pm 140$~km.s$^{-1}$) 
are reached at the edge of our field. The velocity dispersion field exhibits a central
dip with $\sigma_0 \sim 205$~km.s$^{-1}$ and shows the presence
of a cold component (i.e. the inner disc) along the major-axis.

\section{Discussion and conclusions}
\label{sec:conc}

The regularity of the velocity field of the gas suggests that it has settled
onto closed orbits in the centre. In an axisymmetric potential, gas would rapidly
fall onto the equatorial plane of the galaxy in a few orbital periods\footnote{The
orbital time is estimated to be $\sim 26$~Myr at $2\arcsec$.}. This process seems to be
even more efficient if the potential is triaxial and stationary (Colley \& Sparke 1996).
However, tumbling triaxial potentials are known to contain 
stable families of closed orbits which leaves the plane perpendicular to the rotation axis (see e.g.
Magnenat 1982, Mulder \& Hooimeyer 1984). The main family of retrograde orbits is then
the so-called ANomalous Orbits (ANO) which corresponds to the 1:1 vertical resonance.
For slow figure rotation $\Omega_p$ these orbits bifurcates from the E$_z$ family. 
Above a certain critical value $\Omega_{p_{crit}}$, the $z$-orbit become complex 
unstable. The value of $\Omega_{p_{crit}}$
is significantly lowered as soon as a small central mass concentration is present
(see Martinet \& Pfenniger 1987, and Pfenniger \& Friedli 1991 for further details).

We therefore conclude that the potential of NGC~128 must be triaxial and tumbling.
In other words, the prominent peanut in NGC~128 corresponds to a bar viewed nearly edge-on.
The maximum of the peanut corresponds to the axisymmetric horizontal 
and vertical inner Lindblad resonances: $\Omega_p = \Omega - \kappa / 2 =
\Omega - \nu_z / 2$ (Combes et al. 1990). 
In a forthcoming paper, we will present a realistic
mass model of NGC~128 built using the generalization of the Multi-Gaussian
Expansion method (Emsellem 1993). Preliminary results from this model indicate
that the pattern speed must be high with $\Omega_p > 30$~km.s$^{-1}$.kpc$^{-1}$, and therefore
suggests that it overcomes the critical boundary $\Omega_{p_{crit}}$.
The morphology of the retrograde gas orbits in such a case (``fast $\Omega_p$'')
are nicely illustrated in Fig.~3 of Friedli \& Udry (1993):
they are circling the long-axis of the bar, and their edge-on projection shows a central
tilted disc structure very similar indeed to the one observed in NGC~128. 
A detailed dynamical model is required to confirm that this
interpretation is fully consistent with the observed kinematics.

The peanut itself is not perfectly symmetric and exhibits some distortions:
this is certainly the result of the interaction with NGC~127. We suggest
that the interaction with the small satellite NGC~127 (whose integrated luminosity
in the $I_c$ band is $\sim 7\%$ of NGC~128's) triggered the formation of a bar 
in NGC~128 and accelerated its dynamical evolution leading to its peanut-shape.
There is a gap between the two discs along the major-axis which closely 
corresponds to the location of the maximum of the peanut. This is expected if the stars
populating the peanut were driven out of the equatorial plane of the galaxy.
This double disc structures intriguingly ressembles the ones observed in many S0s
(Seifert \& Scorza 1996).

The observed ionized gas in NGC~128 has almost certainly an external origin. 
Our photometric data show that NGC~127 contains a significant 
amount of dust (see Fig.~\ref{fig:hrc}) and therefore very probably 
some gas. It is however difficult to 
say whether this has been accreted to form the tilted red disc observed today in NGC~128:
it could have been formed during an earlier accretion event.
Finally, if NGC~127 is later cannibalized by NGC~128, it is likely that the bar would
be destroyed in the process leading to a nearly axisymmetric bulge (Pfenniger 1991).

\end{document}